\documentclass[sigconf]{acmart}

\AtBeginDocument{%
  }

\copyrightyear{2026}
\acmYear{2026}
\setcopyright{cc}
\setcctype{by}
\acmConference[RecSys '26]{20th ACM Conference on Recommender Systems}{September 27-October 02, 2026}{Minneapolis, MN, USA}
\acmBooktitle{20th ACM Conference on Recommender Systems (RecSys '26), September 27-October 02, 2026, Minneapolis, MN, USA}
\acmDOI{10.1145/3773078.3831874}
\acmISBN{979-8-4007-2284-4/2026/09}

\usepackage{tabularx}
\usepackage{booktabs}
\usepackage{multirow}
\usepackage{subcaption}
\begin{document}

\title{IDProxy: CTR Prediction with Multimodal LLMs for Cold-Start~Recommendation~at~Xiaohongshu}

\author{Yubin Zhang}
\affiliation{%
  \institution{Xiaohongshu Inc.}
  \city{Shanghai}
  \country{China}}
\email{zhangyubin@xiaohongshu.com}

\author{Haiming Xu}
\affiliation{%
  \institution{Xiaohongshu Inc.}
  \city{Shanghai}
  \country{China}}
\email{xuhaiming@xiaohongshu.com}

\author{Guillaume Salha-Galvan}
\affiliation{%
  \institution{Shanghai Jiao Tong University}
  \city{Shanghai}
  \country{China}
}
\email{gsalhagalvan@sjtu.edu.cn}

\author{Ruiyan Han}
\affiliation{%
  \institution{Fudan University}
  \city{Shanghai}
  \country{China}
}
\email{ryhan25@m.fudan.edu.cn}

\author{Feiyang Xiao}
\affiliation{%
  \institution{Fudan University}
  \city{Shanghai}
  \country{China}
}
\email{fyxiao24@m.fudan.edu.cn}

\author{Yanhua Huang}
\affiliation{%
  \institution{Xiaohongshu Inc.}
  \city{Shanghai}
  \country{China}}
\email{yanhuahuang@xiaohongshu.com}

\author{Li Lin}
\affiliation{%
  \institution{Xiaohongshu Inc.}
  \city{Shanghai}
  \country{China}}
\email{lufei2@xiaohongshu.com}

\author{Yang Luo}
\affiliation{%
  \institution{Xiaohongshu Inc.}
  \city{Beijing}
  \country{China}}
\email{yangluo1@xiaohongshu.com}

\author{Yao Hu}
\affiliation{%
  \institution{Xiaohongshu Inc.}
  \city{Beijing}
  \country{China}}
\email{xiahou@xiaohongshu.com}

\renewcommand{\shortauthors}{Yubin Zhang, et al.}

\begin{abstract}
Content-driven platforms such as Xiaohongshu often leverage click-through rate (CTR) prediction models for  recommendation. However, these models depend heavily on item ID embeddings, which perform poorly in item cold-start settings. In this paper, we present IDProxy, a production-scale system developed at Xiaohongshu to address this challenge. IDProxy leverages multimodal large language models (MLLMs) to generate proxy embeddings from rich content signals, enabling CTR prediction for new items in the absence of usage data. Through a lightweight coarse-to-fine mechanism, these proxies are aligned with the ID embedding space and trained end-to-end with the ranking model, allowing seamless integration into production-facing pipelines. Extensive offline and online experiments demonstrate the effectiveness of the method, which has been deployed in 2025 in Xiaohongshu’s Content Feed and Display Ads features, reaching hundreds of millions of users daily.

\end{abstract}



\maketitle

\section{Introduction}
\label{sec:introduction}

Xiaohongshu (RedNote) is a leading content-driven social and e-commerce platform with over 300 million users, engaging with posts and short videos across diverse topics~\cite{xiaohongshu2026}. To manage the massive volume of content, Xiaohongshu relies on large-scale recommendation systems, which are critical for user engagement and retention \cite{zhang2019deep,briand2021semi,li2024recent}, and also deliver personalized advertisements.

Both recommendation and advertising at Xiaohongshu leverage click-through rate (CTR) prediction models to rank items (posts or ads) for each user  \cite{mcmahan2013ad,zhou2018deep,zhang2021deep}. These models predominantly rely on item ID embeddings capturing collaborative signals \cite{zhang2021deep,papadakis2022collaborative}, which nonetheless require sufficient interaction history and thus struggle with new items. This issue, known as the item cold-start problem \cite{zhang2025cold,zhang2025llmtreerec,wang2024large}, is particularly severe on Xiaohongshu, where large volumes of items are continuously uploaded and must be served immediately to ensure a high-quality user experience.


\begin{figure*}[t!]
  \centering
  \includegraphics[width=0.3\linewidth]{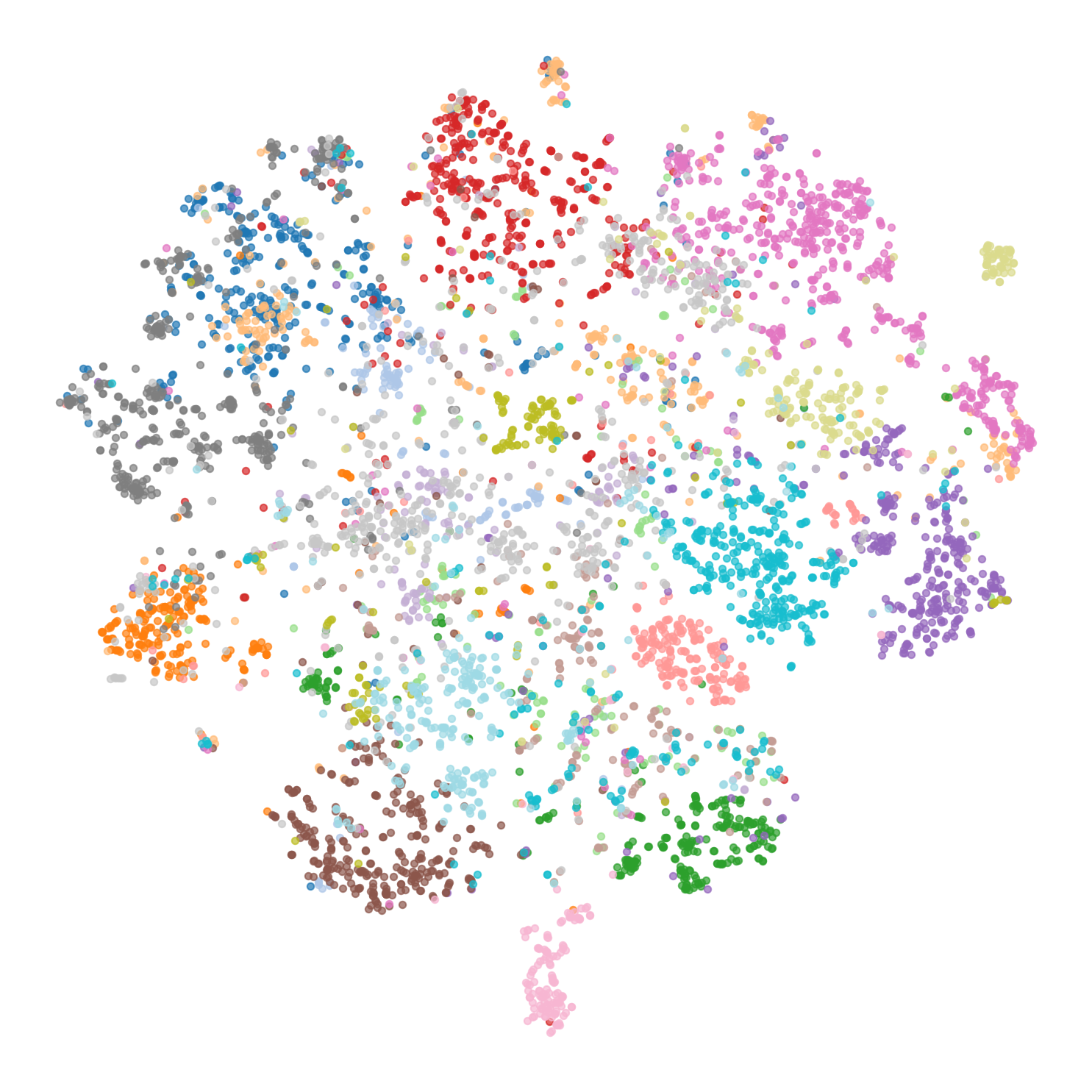}
  \hfill
  \includegraphics[width=0.3\linewidth]{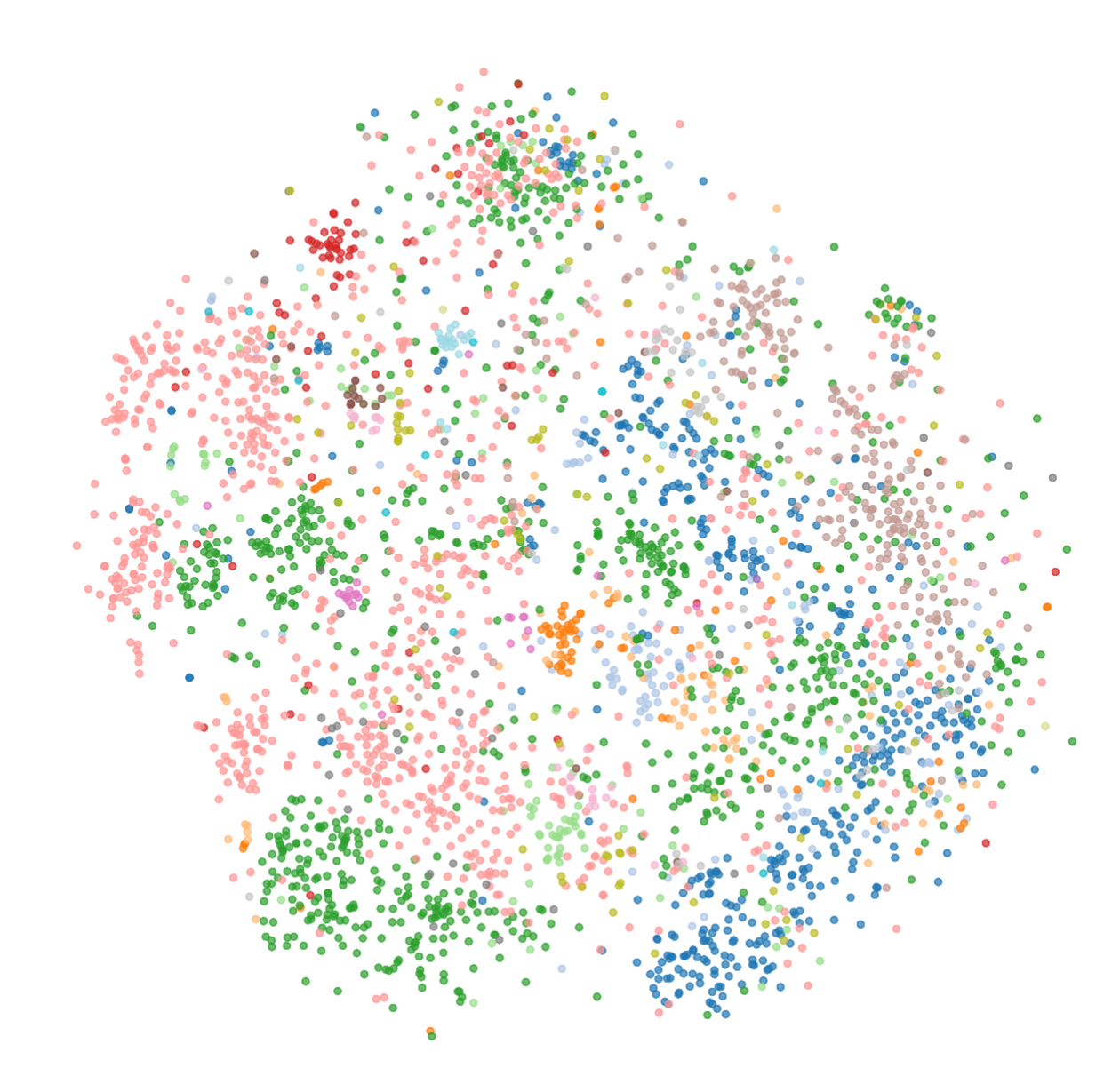}
  \hfill
  \includegraphics[width=0.3\linewidth]{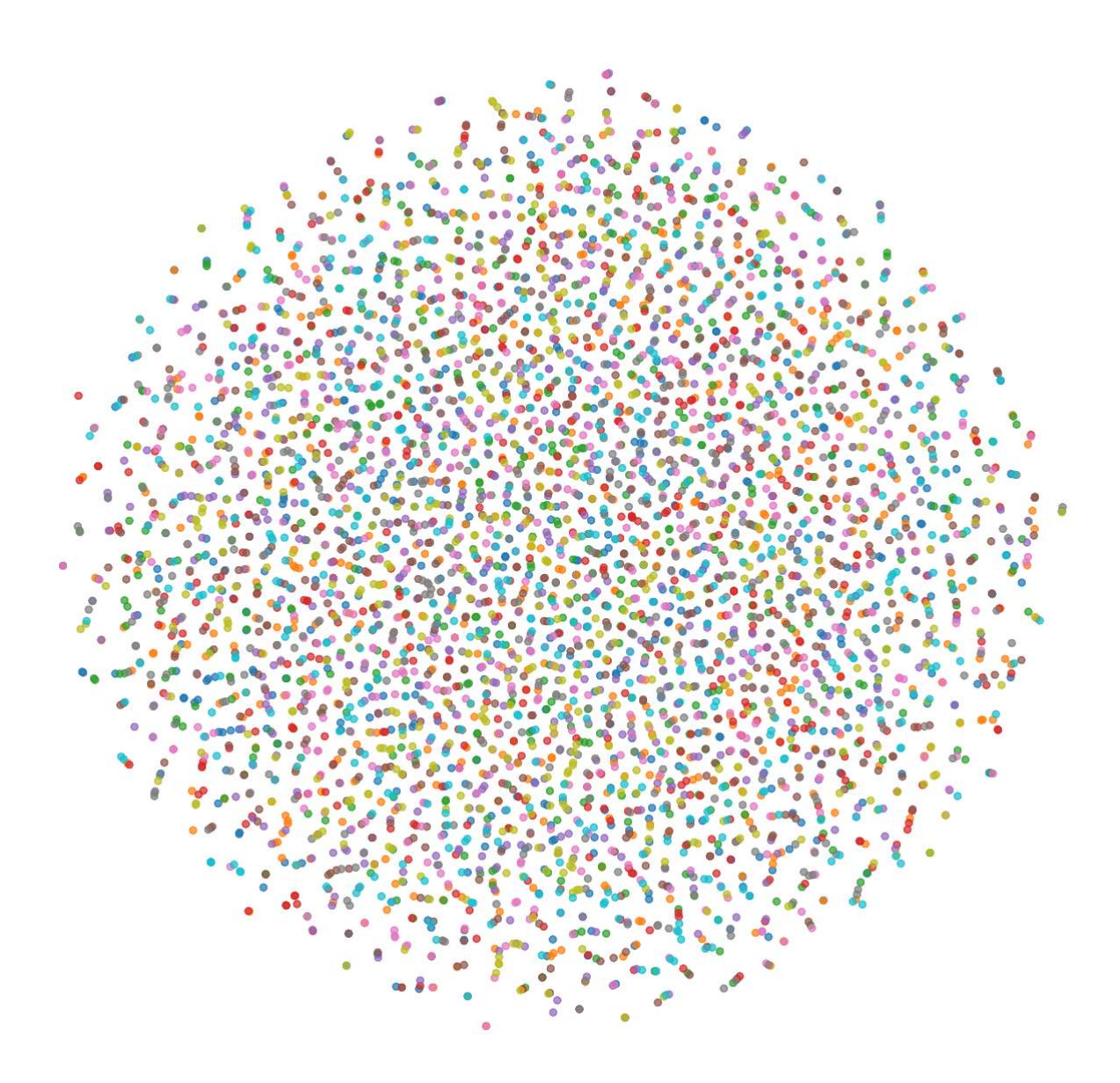}
 \caption{Visualizing item embeddings using t-SNE \cite{maaten2008visualizing}. Left: Multimodal embeddings trained on Swing pairs (e.g., QARM \cite{luo2025qarm}). Middle: MovieLens-1M item ID embeddings learned by SASRec~\cite{SASREC} with feature crossing. Right: Production item ID embeddings from Xiaohongshu. Colors indicate item genres (\textit{"adventure"}, \textit{"sci-fi"},...).}
  \label{fig:visual}
  \Description{cluster visualization of different item ID embedding dataset}
\end{figure*}

Exploiting multimodal content signals such as text and images describing items has emerged as an effective way to learn embeddings without interaction data ~\cite{zhang2025cold,lopez2025survey,pan2022multimodal}. Recent approaches have explored aligning such content representations with collaborative embeddings, either through behavioral co-occurrence patterns ~\cite{luo2025qarm,zhang2026moon} or direct content-to-collaborative mapping ~\cite{barkan_cb2cf_2019,wei2021contrastive}. These methods have achieved promising results, but the former learns embeddings that primarily capture behavioral semantic co-occurrence patterns, which, as shown in Figure 1 (left and right), exhibit significant distributional differences from ID embeddings—revealing inconsistent information representations between the two and limiting their effectiveness as cold-start proxies for ID embeddings. The latter, while validated on academic benchmarks with well-structured ID distributions, similarly struggle with the irregular, non-clustered distributions of industrial ID embeddings, as shown in Figure 1 (middle and right). Moreover, both categories of methods face the challenge of not being able to reuse mature architectures within CTR models and not being fully aligned with the CTR optimization objective. This raises a fundamental question: how can we design an alignment mechanism that bridges multimodal semantic representations and collaborative ID embeddings while maximally reusing existing ID knowledge and structural priors?

This industry paper introduces IDProxy, a production-scale system deployed on Xiaohongshu to address these challenges. IDProxy leverages multimodal large language models (MLLMs)~\cite{ye2025harnessing,chen2024internvl} to generate proxy item embeddings from rich content signals for cold-start CTR prediction. Through a lightweight coarse-to-fine mechanism, these proxies are aligned and optimized end-to-end under CTR objectives jointly with the ranking model, enabling seamless reuse of its architecture and continual evolution in production pipelines. We demonstrate the effectiveness of IDProxy through offline experiments and online A/B tests. IDProxy now powers both Content Feed and Display Ads on Xiaohongshu’s Explore Feed, serving hundreds of millions of users daily.

This paper is organized as follows. Section~\ref{sec:preliminaries} formally introduces the problem and reviews related work. Section~\ref{sec:method} presents IDProxy. Section~\ref{sec:experiments} reports experimental results, and Section~\ref{sec:conclusion} concludes.

\section{Preliminaries}
\label{sec:preliminaries}

We begin this section with a formal definition of the problem addressed in this paper, followed by a discussion of related work.

\subsection{Problem Formulation}
\label{subsec:problemformulation}

We consider a set $\mathcal{U}$ of users and a set $\mathcal{I}$ of items on an online service. Each user-item sample is represented as a tuple $(u, i, \mathbf{x}_{ui})$, where $u \in \mathcal{U}$ is the user ID, $i \in \mathcal{I}$ is the item ID, and $\mathbf{x}_{ui}$ denotes contextual features such as user profiles, item attributes, and interaction history.
For each sample, a CTR model $f_{\theta}$ predicts the probability that user $u$ clicks on item $i$ as follows: \begin{equation}
\hat{y}_{ui} = f_{\theta}(\mathbf{e}_u, \mathbf{e}_i, \mathbf{x}_{ui}) \in [0,1],
\end{equation}
where $\mathbf{e}_u \in \mathbb{R}^{d}$ and $\mathbf{e}_i \in \mathbb{R}^{d}$ denote learnable $d$-dimensional user and item ID embeddings, respectively, and $\theta$ denotes the model's parameters. These predictions are then used to rank items for recommendation.

We focus on the cold-start setting, where new items lack sufficient interaction history, leading to poorly trained item ID embeddings $\mathbf{e}_i$ \cite{zhang2025cold,zhou2023contrastive}. However, these items are associated with rich multimodal content, such as images and textual descriptions. Our goal is to exploit this content to learn item proxy embeddings $\mathbf{p}_i \in \mathbb{R}^{d}$ that encode informative semantics and are well aligned with the distribution of existing item ID embeddings, such that $\mathbf{p}_i$ can be seamlessly substituted for $\mathbf{e}_i$ in the existing CTR model.

\subsection{Related Work}
\label{subsec:relatedwork}

Cold-start issues have been extensively studied over the years~\cite{schein2002methods,van2013deep,he2016vbpr,briand2021semi,salha2021cold,jangid2025deep,smith2017two,wang2018billion,volkovs2017dropoutnet,zhou2023contrastive,wang2024large,huang2025large,xu2022alleviating,pan2019warm}. We refer to recent surveys~\cite{zhang2025cold,li2024recent,zhang2019deep,jangid2025deep} for comprehensive reviews and focus here on work most closely related to our problem formulation. Aligning multimodal content representations with collaborative embeddings for CTR-based ranking remains challenging. Recent methods such as QARM \cite{luo2025qarm}, MOON \cite{zhang2026moon}, and SimTier\&Maker~\cite{sheng_maker_2024} align co-occurrence structures across modalities, followed by dense feature quantization or auxiliary CTR-related training. While effective in some settings, these approaches rely on manually designed alignment objectives and often require substantial experimentation to ensure compatibility with existing CTR models. Moreover, they do not fully exploit the architectures and distributional properties of pre-existing ranking models, resulting in limited gains and increased deployment complexity.

Another line of work maps content directly to collaborative embeddings, typically using multi-layer perceptrons (MLPs). Representative examples include CB2CF~\cite{barkan_cb2cf_2019} and CLCRec~\cite{wei2021contrastive}, which employ regression or contrastive objectives, as well as generative approaches such as GoRec~\cite{bai2023gorec} and GAR~\cite{chen2022generative}. However, these methods face limitations in industrial CTR systems. As shown in Figure~\ref{fig:visual}, ID embeddings learned on public benchmarks such as MovieLens~\cite{movielen} exhibit clear cluster-like semantic structures, making alignment with content features comparatively easier. In contrast, industrial ID embeddings often exhibit irregular, non-clustered distributions due to feature sparsity and complex interaction patterns~\cite{zhang2022towards}. In such settings, frozen content encoders, or shallow MLP mappings are often insufficient to bridge heterogeneous spaces. More broadly, generating item ID proxies that remain aligned with evolving real-world CTR models raises additional challenges, including learning against static targets in continuously updated systems and failing to reuse the ranking model’s evolving structure, which limits long-term optimization and stable deployment. To address this, we propose a two-stage alignment approach that first coarsely aligns content features with the collaborative space using MLLM, then refines the alignment through end-to-end optimization with the CTR ranking loss.

Finally, recent advances in large language models (LLMs) and multimodal LLMs open new opportunities for cold-start recommendation \cite{wu2024survey,zhang2025cold,li2024large,zhang2025llmtreerec,wang2024large,huang2025large,zhang2025metric}. Beyond final-layer outputs, intermediate hidden representations have been shown to capture richer and more transferable semantic information \cite{liu2019linguistic, shwartz2017opening,skean2025layer}, generalize better to downstream tasks \cite{skean2024does, skean2025layer} and even sometimes benefit from discarding deeper layers \cite{bordesguillotine}. These findings motivate leveraging multi-layer MLLM hidden representations to extract finer-grained content features for cold-start recommendation. 

\section{IDProxy: Proxy ID Embeddings with~MLLMs}
\label{sec:method}


\begin{figure*}[t]
    \centering
    \includegraphics[width=\linewidth]{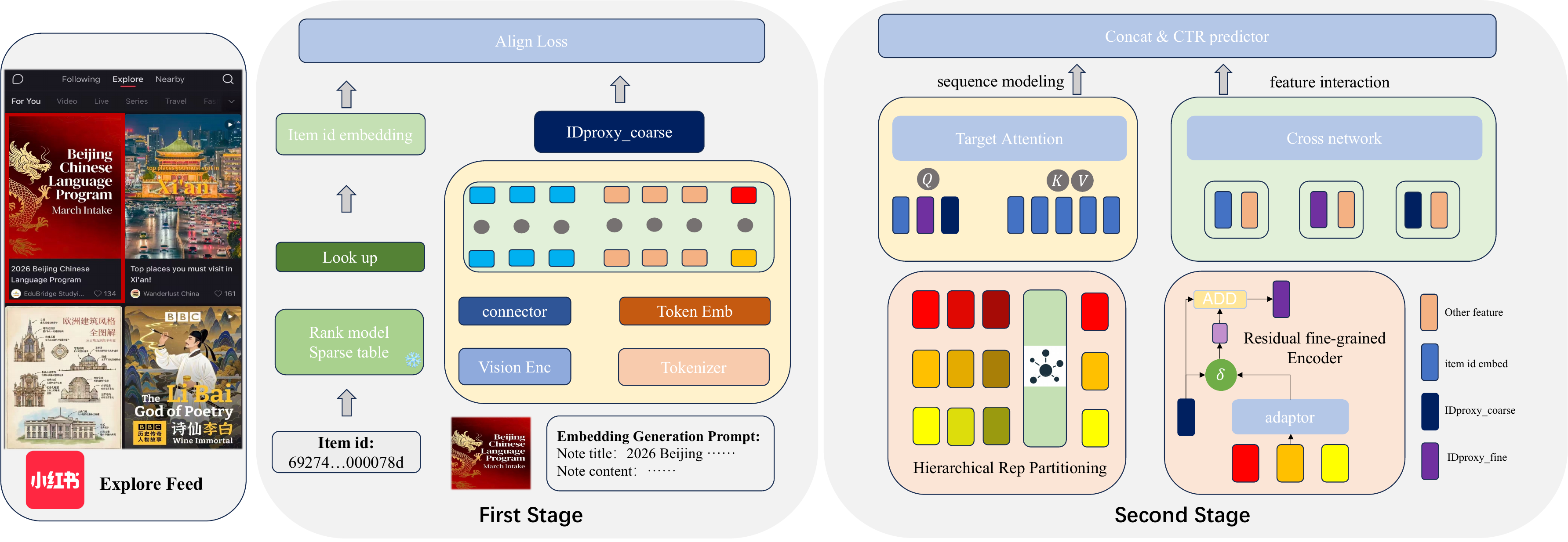}
  \caption{Overview of IDProxy's two-stage coarse-to-fine alignment framework, used on Xiaohongshu's Explore Feed (left).}
  \label{fig:framework}
  \Description{The overall framework of out method}
\end{figure*} 

This section introduces IDProxy which, as illustrated in Figure~\ref{fig:framework}, consists of (i) an MLLM-based coarse proxy generation stage and (ii) an end-to-end alignment stage that refines these proxies using MLLM hidden states jointly with the CTR ranker, all while reusing its core architecture. This coarse-to-fine paradigm aligns proxy and item ID embeddings, mitigating distributional discrepancies, and enabling reuse of the ranker's feature and structural~priors.

\subsection{Stage 1: MLLM-based Proxy Generation}


Rather than simple MLP mappings, Stage~1 leverages an MLLM to learn proxy embeddings for cold-start items, approximating the CTR model's ID embedding distribution via contrastive learning.

\subsubsection{ID Embedding Preprocessing}
Let $\mathbf{e}_i^{\text{raw}} \in \mathbb{R}^d$ denote the ID embedding of item $i \in \mathcal{I}$ learned by the online CTR model. As many items receive limited interactions in industrial systems~\cite{wang2018billion}, we apply a frequency threshold $\tau$ to filter out items updated fewer than $\tau$ times, ensuring reliable alignment targets. Moreover, since embedding magnitude often correlates with item popularity, we apply $\ell_2$ normalization to remove magnitude-induced bias and stabilize alignment learning~\cite{kim_test_2023}:
\begin{equation}
\mathbf{e}_i = \frac{\mathbf{e}_i^{\text{raw}}}{\|\mathbf{e}_i^{\text{raw}}\|_2} \in \mathbb{R}^d.
\end{equation}

\subsubsection{MLLM-Based Multimodal Encoding}
Each item $i$ has multimodal content, including text and images. We use an MLLM $M$ as the content encoder with the following input for proxy generation:

\noindent\fbox{%
    \parbox{\dimexpr\linewidth-2\fboxsep-2\fboxrule\relax}{%
        \textbf{Prompt:} [BOS]\textless image\textgreater \textless text\textgreater The compression
        word is:"[EMB]". [EOS]%
    }%
}
[BOS], [EOS] and [EMB] are special tokens, while \textless image\textgreater~and \textless text\textgreater~are placeholders replaced with content.

We use the final token-level hidden states $\mathbf{H}_i \in \mathbb{R}^{T \times D}$ to construct a content embedding $\mathbf{z}_i = g(\mathbf{H}_i) \in \mathbb{R}^D$ for item $i$, where $g(\cdot)$ is an attention-based aggregation over tokens, including  [EMB]. Then, $\mathbf{z}_i$ is projected into the ID embedding space by an MLP $\phi$ followed by $\ell_2$ normalization: 
\begin{equation}
\tilde{\mathbf{h}}_i = \frac{\phi(\mathbf{z}_i)} {\|\phi(\mathbf{z}_i)\|_2} \in \mathbb{R}^d.
\end{equation}



\subsubsection{Proxy Alignment}
For training, we adopt a contrastive objective that pulls $\tilde{\mathbf{h}}_i$ toward its corresponding ID embedding $\mathbf{e}_i$, while pushing it away from other item embeddings. Given a mini-batch $\mathcal{B}$, we treat $(\tilde{\mathbf{h}}_i, \mathbf{e}_i)$ as a positive pair and $(\tilde{\mathbf{h}}_i, \mathbf{e}_j)$ for $j \neq i$ as in-batch negatives. We define the Proxy Alignment Loss $\mathcal{L}_{\mathrm{PAL}}$ as:
\begin{equation} \mathcal{L}_{\text{PAL}} = - \frac{1}{|\mathcal{B}|} \sum_{i \in \mathcal{B}} \log \frac{\exp\left(\tilde{\mathbf{h}}_i^\top \mathbf{e}_i / \tau_c \right)} 
{\sum_{j \in \mathcal{B}} \exp\left(\tilde{\mathbf{h}}_i^\top \mathbf{e}_j / \tau_c \right)}, \end{equation} 
for some temperature $\tau_c > 0$. Parameters of $M$ and $\phi$ are jointly optimized via gradient descent to minimize $\mathcal{L}_{\text{PAL}}$. After convergence, we define the coarse proxy embedding for item $i$ as $\mathbf{p}_i^{\text{coarse}} = \tilde{\mathbf{h}}_i$.


\subsection{Stage 2: CTR-Aware Proxy Alignment}

While Stage 1 proxies lie in the static item ID embedding space, coarse alignment alone cannot fully exploit the end-to-end optimization and evolving structural priors of CTR models. Motivated by evidence  that MLLM hidden states encode rich semantics (see Section~\ref{subsec:relatedwork}), we add a fine-grained alignment stage that leverages these states to refine proxies via end-to-end training with the CTR model using a lightweight adaptor.




\subsubsection{Hierarchical Representation Partitioning}

We extract hidden states from multiple layers of the MLLM $M$ trained in Stage~1, which typically consists of several tens of layers. To balance representational richness and learning efficiency, we use $k$-means clustering \cite{lloyd1982least} to extract three layer subgroups $l_{n_1}$, $l_{n_2}$, and $l_{n_3}$, capturing hierarchical information from shallow to deep representations. We then apply the Stage~1 pooling function $g(\cdot)$ to each subgroup:
\begin{equation}
\mathbf{z}_i^{(l)} = g(\mathbf{H}_i^{(l)}) \in \mathbb{R}^D, \text{ for } l \in \{l_{n_1}, l_{n_2}, l_{n_3}\}.
\end{equation}


\subsubsection{Lightweight Multi-Granularity Adaptor for CTR}

The hidden states are already partially aligned with item ID embeddings via Stage~1. We refine them to capture fine-grained CTR-relevant information using a lightweight and multi-granularity  adaptor:
\begin{equation}
\mathbf{p}_i^{\text{raw\_fine}} =
\tilde{\phi} (\text{Concat} (\mathbf{z}_i^{(l_{n_1})}, \mathbf{z}_i^{(l_{n_2})}, \mathbf{z}_i^{(l_{n_3})}))
\in \mathbb{R}^{\tilde{d}},
\end{equation}
where $\tilde{\phi}$ is an MLP with far fewer parameters than the MLLM or ranking model, fusing multi-granularity representations into the ranker embedding space.
Since $\mathbf{p}_i^{\text{raw\_fine}}$ may contain information redundant with $\mathbf{p}_i^{\text{coarse}}$, we further introduce a residual gating module to adaptively control the contribution of fine-grained signals, encouraging the model to capture information beyond the coarse proxy. The final item proxy embedding, which fuses coarse- and fine-grained representations,~is:
\begin{equation}
\mathbf{p}_i^{\text{fine}} = W_c \mathbf{p}_i^{\text{coarse}} + r \odot \mathbf{p}_i^{\text{raw\_fine}},\end{equation}
with $r = \sigma (W_g [\mathbf{p}_i^{\text{coarse}}, \mathbf{p}_i^{\text{raw\_fine}}])$.



\subsubsection{End-to-End Joint Training with CTR Ranker}

In modern CTR ranking models, feature interaction and sequence modeling are critical components (see Figure~\ref{fig:framework}). To inject fine-grained item information into these components, we augment the original feature set, composed of the item ID and $\mathbf{p}_i^{\text{coarse}}$, with the refined proxy $\mathbf{p}_i^{\text{fine}}$, for both the feature interaction and target attention modules.

Let $f_{\theta}$ denote the CTR ranking model, with parameters $\theta$. For a training instance $(u, i, \mathbf{x}_{ui}, y_{ui})$, with a click/no-click signal $y_{ui} \in \{0,1\}$, the CTR prediction becomes:
\begin{equation}
\hat{y}_{ui} = f_\theta(\mathbf{e}_u, \mathbf{e}_i, \mathbf{p}_i^{\text{coarse}}, \mathbf{p}_i^{\text{fine}}, \mathbf{x}_{ui}).
\end{equation}
$f_{\theta}$ is trained for classification, e.g., using a cross-entropy loss:
\begin{equation} \mathcal{L}_{\text{CTR}} = - \frac{1}{|\mathcal{D}|} \sum_{(u, i, \mathbf{x}_{ui}, y_{ui}) \in \mathcal{D}} [ y_{ui} \log \hat{y}_{ui} + (1-y_{ui}) \log (1-\hat{y}_{ui})]. 
\end{equation}
During Stage~2, parameters of the MLP adaptor $\tilde{\phi}$, together with $W_g$, $W_c$, and $\theta$, are optimized via gradient descent, while MLLM parameters are frozen. This end-to-end optimization under the CTR objective enables $\mathbf{p}_i^{\text{fine}}$ to capture information beyond $\mathbf{p}_i^{\text{coarse}}$ and allows seamless integration into the CTR production system.

\subsubsection{Deployment} We stress that IDProxy is designed for large-scale deployment. The lightweight multi-granularity adaptor only needs to be trained offline once, after which its parameters can be stored and packaged together with the MLLM as an IDProxy generation service. In practice, for each new item, we compute coarse and fine proxies in real time and write them into online storage. Production-facing recommender systems can then retrieve them via IDs for seamless integration into the CTR ranking pipeline.

\section{Experimental Analysis}
\label{sec:experiments}
We evaluate IDProxy through large-scale offline experiments and online A/B tests, designed to address the following questions:
\begin{itemize}
    \item \textbf{Q1}: What is the contribution of the coarse stage, and how does IDProxy compare with existing alignment methods?
    \item \textbf{Q2}: What is the contribution of the fine stage, i.e., what is the impact of end-to-end modeling and structural reuse?
    \item \textbf{Q3}: Does IDProxy effectively alleviate the cold-start problem for newly published content on Xiaohongshu?
    \item \textbf{Q4}: What is the impact of IDProxy on real-world business metrics in large-scale industrial scenarios at Xiaohongshu? 
\end{itemize}

\subsection{Experimental Setup}

We first conduct offline CTR prediction experiments on historical click logs from 2025. We compare IDProxy to the production baseline at Xiaohongshu, denoted as Base. We adopt InternVL~\cite{chen2024internvl} as the MLLM and train all methods using the AdamW optimizer~\cite{loshchilov2017decoupled} with a learning rate of $1\times10^{-4}$ and a batch size of 512. Base is a highly optimized industrial CTR system incorporating complex ID-based feature interactions and user sequential behavior modeling. For both methods, some implementation details are omitted for confidentiality reasons. In our system, an absolute increase of 0.0010 in AUC is typically regarded as practically significant.

We also run online A/B tests on millions of users in Xiaohongshu's Explore Feed for content discovery, deploying IDProxy to a randomly selected treatment group in two scenarios: Content Feed, which recommends user posts (test done in August 2025), and Display Ads, which serves ads (in March 2025). For Content Feed, we report Time Spent, number of Reads (clicks), and number of Engagements (including likes, comments...) as evaluation metrics. For Display Ads, we report standard ads metrics: Advertiser Value (ADVV) and COST to measure business value, as well as Impression and CTR to reflect user experience.

\subsection{Results and Discussion}

\subsubsection{Offline Performance of Stage~1 (Q1)}
The first four rows of Table~\ref{tab:offline_results} compare Stage~1 of IDProxy (v3) with multimodal baselines. Mainstream multimodal embeddings (v1, similar to NoteLLM2 \cite{zhang2025notellm}) yield limited improvements over Base due to misalignment with the item ID embedding distribution. Similarly, using frozen embeddings with an MLP mapper (v2, CB2CF-style) achieves only marginal gains, which confirms the limitations of simple MLP-based mappings in bridging the semantic gap between content representations and the irregular distribution of industrial ID embeddings. By explicitly fitting the ID distribution and leveraging an MLLM for alignment, Stage~1 of IDProxy (v3) outperforms both approaches.


\begin{table}[t]
\centering
\caption{Offline CTR prediction gains relative to the baseline.} \label{tab:offline_results} 
\resizebox{\linewidth}{!}{
\begin{tabular}{l|c|c} 
\toprule 
\textbf{Model Variant} & \textbf{Model ID} & \textbf{$\Delta$ AUC} \\ \midrule 
Base (Production Baseline at Xiaohongshu) & - & 0 \\
Base + NoteLLM2-Like Embed  & v1 & $+ 0.015\%$ \\
Base + Static Vector (MLP Mapping) & v2 & $+ 0.02\%$ \\ 
Base + IDProxy (Stage 1) & v3 & $+0.05 \%$ \\
\midrule
Base + IDProxy (Stage 1 + 2, w/o Structure Reuse) & v4 & $ + 0.08\%$ \\ 
\textbf{Base + IDProxy (Stage 1 + 2)} & \textbf{v5} & $\textbf{+0.14\%}$ \\ \bottomrule
\end{tabular}}
\end{table}

\begin{table}[t]
\centering
\caption{Online $\Delta$AUC gains for global notes and new notes.}
\resizebox{\linewidth}{!}{
\begin{tabular}{l|ccccc}
\hline
 & \textbf{Day 1} & \textbf{Day 2} & \textbf{Day 3} & \textbf{Day 4} & \textbf{Day 5} \\ \hline
\textbf{Global Notes} & $+0.13\%$ & $+0.15\%$ & $+0.14\%$ & $+0.12\%$ & $+0.15\%$ \\
\textbf{New Notes}    & $\textbf{+0.24\%}$ &$ \textbf{+0.32\%}$ & $\textbf{+0.23\%}$ & $\textbf{+0.27\%}$& $\textbf{+0.31\%}$ \\ \hline
\end{tabular}}
\label{tab:online_new_item}
\end{table}

\begin{table}[t]
\centering 
\caption{Online A/B tests on Xiaohongshu's Explore Feed.} 
\resizebox{\linewidth}{!}{
\begin{tabular}{c|c|c|c|c}
\toprule
\textbf{Scenario}         & \multicolumn{4}{c}{\textbf{Online A/B Test Metric}} \\ \midrule
\multirow{2}{*}{Using IDProxy in \textbf{Content Feed}} & \textbf{Time Spent}    & \textbf{Reads}   & \textbf{Engagements} & -     \\  
    & $+0.22\%$ & $+0.39\%$ & $+0.5\%$      & - \\ \midrule       
    \multirow{2}{*}{Using IDProxy in \textbf{Display Ads}} & \textbf{Impression}  & \textbf{ADVV}  & \textbf{COST}  & \textbf{CTR} \\ 
    & $+1.28\% $  & $+1.93\%$    & $+1.73\%$   & $+0.23\% $ \\ \bottomrule
\end{tabular}}
\label{tab:online_abtest} 
\end{table}

\subsubsection{Offline Performance of Stage2 (Q2)}
The last two rows of Table~\ref{tab:offline_results} evaluate the effect of adding Stage~2. In v4, MLLM hidden features are treated as ordinary item features, concatenated with existing inputs and trained end-to-end using the CTR loss. This already improves performance over v3, indicating the benefit of end-to-end learning. However, v4 does not explicitly reuse the Base model’s sequence and feature interaction structures. In contrast, v5 integrates IDProxy directly into the ranker’s atomic ID slots, leading to a more substantial performance gain of $+0.14\%$. This confirms the benefit of allowing multimodal features to inherit the well-trained structural priors of ID-based CTR models. Importantly, we note that although the $\Delta$AUC may appear small, these results were considered valuable in practice within the company, given the strength of the Xiaohongshu production Base model.

\subsubsection{Online Cold-Start (Q3) and Business (Q4) Performance}

Table~\ref{tab:online_new_item} compares the online AUC performance of IDProxy over a 5-day period for both global traffic and "new notes", i.e., posts published within 24 hours. The AUC lift for global notes stabilizes around $0.12\%$–$0.15\%$, whereas new notes exhibit substantially larger gains of $0.23\%$–$0.32\%$. This roughly $2\times$ improvement demonstrates that IDProxy effectively transfers semantic information to items lacking interaction history, boosting cold-start performance. In addition,
Table~\ref{tab:online_abtest} reports statistically significant improvements (at the 1\% level) across all online metrics in both Content Feed and Display Ads scenarios. Overall, these consistent results confirm the robustness and business impact of IDProxy in large-scale industrial deployment.

\subsubsection{Limitations and Future Work} 
Finally, while the results demonstrate promising performance, we acknowledge that IDProxy has some limitations that motivate several interesting avenues for future research. Although we considered a wide range of settings, the alignment of sparse ID features still relies entirely on contrastive loss with in-batch negatives, without a systematic investigation of alternative alignment objectives. Future work will explore more principled objectives beyond contrastive loss—such as optimal transport-based distribution matching or regression-based approaches \cite{meng2020wasserstein}—which may better capture the geometry of the industrial ID embedding space.
Moreover, the MLLM is optimized to align with static ID embeddings. Looking ahead, we plan to investigate end-to-end joint optimization of the MLLM and CTR model via parameter-efficient methods such as LoRA \cite{hu2022lora}. This would enable direct CTR-gradient flow into the content encoder, allowing the model to learn task-specific representations without being bottlenecked by the intermediate ID embedding space.

\section{Conclusion}
\label{sec:conclusion}
This industry paper introduced IDProxy, a large-scale system that leverages MLLMs to address cold-start CTR prediction in content recommendation. Following the successful A/B tests presented in this work, IDProxy was deployed in late 2025 to all users and now powers both Content Feed and Display Ads at Xiaohongshu, serving hundreds of millions of users daily. Beyond its industrial impact, our study demonstrates the benefits of MLLM-based alignment of multimodal content with item ID embeddings, and the importance of end-to-end refinement with CTR models while reusing their core components. Our results also confirm the potential of MLLM hidden states for content representations, motivating future research on their integration for recommendation.

\section*{Speaker Bios}
\textbf{Yubin Zhang} is a Researcher at Xiaohongshu. As a member of the Department of Algorithms, he develops large-scale machine learning and LLM systems to improve content recommendation and discovery for hundreds of millions of users on the platform. He received his Master's degree from Tongji University in China.
\textbf{Guillaume Salha-Galvan} is an Associate Professor at Shanghai Jiao Tong University (SJTU). As a faculty member of the SJTU Paris Elite Institute of Technology, he conducts research on LLMs and their applications to recommender systems. He regularly serves as a reviewer at RecSys and has co-authored nine RecSys papers over the past six editions, including nominations for Best Short Paper (2020), Best Student Paper (2021), and Best Full Paper (2024). He received his Ph.D. degree from École Polytechnique in France.





\bibliographystyle{ACM-Reference-Format}
\bibliography{sample-base}


\end{document}